\documentclass[%
notitlepage,
12pt,
 amsmath,amssymb,
aps,
prstab,
superscriptaddress,
]{revtex4-2}

\usepackage{times}
\usepackage{amsmath}
\usepackage{amssymb}
\usepackage{amsfonts}
\usepackage{amsthm}
\usepackage{float}
\usepackage{bm}
\usepackage{latexsym}
\usepackage{hyperref}
\usepackage{multirow}
\usepackage[capitalize]{cleveref}
\usepackage{graphicx}
\hypersetup{
    colorlinks,
    citecolor=blue,    filecolor=blue,
    linkcolor=blue,    urlcolor=blue
}
\usepackage{lineno}

\begin{document}


\title{Compact dose delivery of laser-accelerated high-energy electron beams \\ towards radiotherapy applications }

\author{Bing Zhou}
 \affiliation{Laboratory of Zhongyuan Light, School of Physics, Zhengzhou University, Zhengzhou, 450001, China}
 \affiliation{Department of Engineering Physics, Tsinghua University, Beijing , China}
 
\author{Zhiyuan Guo}
 \affiliation{Department of Engineering Physics, Tsinghua University, Beijing , China}
 
\author{Yang Wan}
 \email{yangwan23@zzu.edu.cn}
 \affiliation{Laboratory of Zhongyuan Light, School of Physics, Zhengzhou University, Zhengzhou, 450001, China}
 \affiliation{Beijing Academy of Artificial Intelligence, Beijing, China}

\author{Shuang Liu}
 \affiliation{Department of Engineering Physics, Tsinghua University, Beijing , China}
 
\author{Jianfei Hua}
 \affiliation{Department of Engineering Physics, Tsinghua University, Beijing , China}
 
\author{Wei Lu}
\email{weilu@ihep.ac.cn}
\affiliation{Institute of High Energy Physics, Chinese Academy of Sciences, Beijing 100049, China}
\affiliation{Department of Engineering Physics, Tsinghua University, Beijing , China}


\begin{abstract}
The use of very high energy electron (VHEE) beams for radiotherapy has been actively studied for over two decades due to their advantageous dose distribution, deep penetration depth and great potential of ultra-high dose-rate irradiation. Recently, laser-plasma wakefield accelerator (LWFA) has emerged as a promising method for the compact generation of VHEE beams, due to its substantially higher accelerating gradients compared to traditional radio-frequency accelerators. However, how to compactly deliver the LWFA-based VHEE beams of relatively large energy spread and create a maximum dose deeply inside the body remains very challenging. In this article, we present a simple dose delivery scheme utilizing only two dipole magnets for LWFA-based VHEE treatment. By adjusting the magnet strengths, the electron beams can be guided along different angular trajectories towards a precise position as deep as 20 cm within a water phantom, creating a maximum dose over the target region and significantly reducing the entrance dose. Supported by Monte Carlo simulations, such a beam delivery approach is demonstrated to be insensitive to the beam energy spread and meanwhile capable of controlling precisely the dose-peak position in both lateral and longitudinal directions. As such, a uniform dose peak can be generated by the weighted sum of VHEE beams that reach different dose-peak depths. These results demonstrate that LWFA-based VHEE beams can be compactly delivered into a deep-seated tumor region in a controllable manner, thus advancing the development of the VHEE radiotherapy towards the practical clinical applications in the near future. 
\end{abstract}

\maketitle


\section{\label{sec:level1}Introduction}


Radiotherapy is widely applied in oncology by employing ionizing radiation to either suppress or control the growth of tumor tissue. Currently, three radiation sources are mainly used in radiotherapy: photons, ions and low energy electrons (less than $20$ MeV). Among them, photon therapy offers significant penetration depth but results in high dose deposition in normal tissues. To reduce this risk, high conformal techniques such as intensity-modulated radiation therapy (IMRT)\cite{bentzen2005radiation} and volumetric modulated arc therapy (VMAT) \cite{elith2011introduction, otto2008volumetric} were introduced. hadron  therapy, on the other hand, leverages the Bragg peak effect of protons and heavy ions, which allows for substantial energy deposition at the range’s end while reducing the entrance dose. However, hadron treatment are quite sensitive to biological matter inhomogeneities, with minor disturbances causing shifts in Bragg peak localization\cite{urie1984compensating}, and also requires substantial infrastructure and high costs. Conventional electron beam therapy primarily employs low-energy electron beams, which are suitable for treating surface or shallow tumors because of their high entrance dose and the rapid decrease in dose with depth.

Recently, very high-energy electron beams ($50-250$ MeV) have been proposed as an alternative candidate for radiotherapy\cite{desrosiers2000150, ronga2021back}. Compared to low-energy electrons, VHEE can penetrate much deeper region with sharper lateral penumbra profiles\cite{desrosiers2000150}. Meanwhile, several researches have shown that they exhibit superior therapeutic efficacy to photon modalities for several different tumours \cite{yeboah2002optimized,fuchs2009treatment,bazalova2015treatment,schuler2017very,zhang2023treatment} and less sensitive to inhomogeneities than ion beams\cite{desrosiers2000150,lagzda2020influence}. However, the accelerating gradient of conventional radio-frequency (RF) accelerators is generally much less than 100 MeV/m, which necessitates a linear accelerator of at least tens of meters in length to produce such very high-energy electron beams, making them difficult to be adapted into a common treatment room \cite{Wuensch2021facility}. 

Over the past two decades, strong plasma wave driven by ultrashort ultraintense lasers for accelerating electron beams, known as laser wakefield acceleration (LWFA) \cite{Tjima1979accelerator,esarey2009physics} has attracted great interest worldwide, due to its ultrahigh accelerating gradient of above 100 GeV/m, more than 1000 times stronger than the RF accelerators. With the rapid development of LWFA technology, this new kind of accelerator is now capable of stably delivering VHEE beams of several hundreds of MeV \cite{geddes2004high,mangles2004monoenergetic,faure2004laser} with small source sizes (few microns scale) \cite{weingartner2012ultralow,wan2023femtosecond} and low divergence (few milliradians scale) \cite{albert20212020,PhysRevX.10.031039} within only few millimeters distance, thereby making VHEE radiotherapy platforms much more compact and economically viable \cite{malka2008principles,joshi2020perspectives}. Recently, several research groups have focused on the studies of LWFA-based VHEE radiotherapy, including the measurements of dose distribution \cite{glinec2006radiotherapy,lundh2012comparison,Svendsen2021LPA-VHEE}, treatment planning \cite{fuchs2009treatment} and irradiating mice tumors with a prototype LWFA machine \cite{guo2025preclinical}. However, how to transport such a VHEE beam generated from LWFA to target the deeply-seated tumor precisely remains challenging due to the following two aspects. Firstly, VHEEs normally cannot create a dose peak at the tumour target similar as ion beams, but exhibit high dose depositions at both entry and exit points, which could possibly damage the surface skin and other healthy issues \cite{desrosiers2000150}. Additionally, VHEEs from laser wakefield accelerators normally have a notable RMS energy spread of over 5\% if ensuring high charge (more than 100 pC) \cite{esarey2009physics,joshi2020perspectives}, significantly complicating beam transport modalities.

Previous studies have explored the use of quadrupole magnets to focus the VHEE beams at a specific depth within a water phantom\cite{glinec2006radiotherapy,whitmore2021focused,whitmore2023cern,reaz2022sharp,kokurewicz2021experimental,Svendsen2021LPA-VHEE}. However, quadrupole magnets are not suitable for beams with large energy dispersion, such as LWFA-based ones. Moreover, current experimental findings do not support the deposition of peak dose beyond a depth of 6 cm in water phantoms \cite{kokurewicz2021experimental,whitmore2023cern}. In this article, we propose a simple approach to effectively deliver the peak dose of VHEE into the tumor target by employing only two dipole magnets. These magnets can guide the electron beams from different directions to achieve a maximum dose at a depth of up to 20 cm inside the water phantom. Monte Carlo simulations also demonstrate that this technique exhibits exceptional resistance to beam energy spread, particularly attractive for LWFA-based applications. Additionally, it allows precise control of the dose peak position both laterally and longitudinally. These advancements afford promises for the development of a compact VHEE radiotherapy machine with high precision. 

\section{Results and Discussions}

\subsection{The basic concept of the proposed VHEE beam delivery system}

\begin{figure*}
    \centering
    \includegraphics[width=1\linewidth]{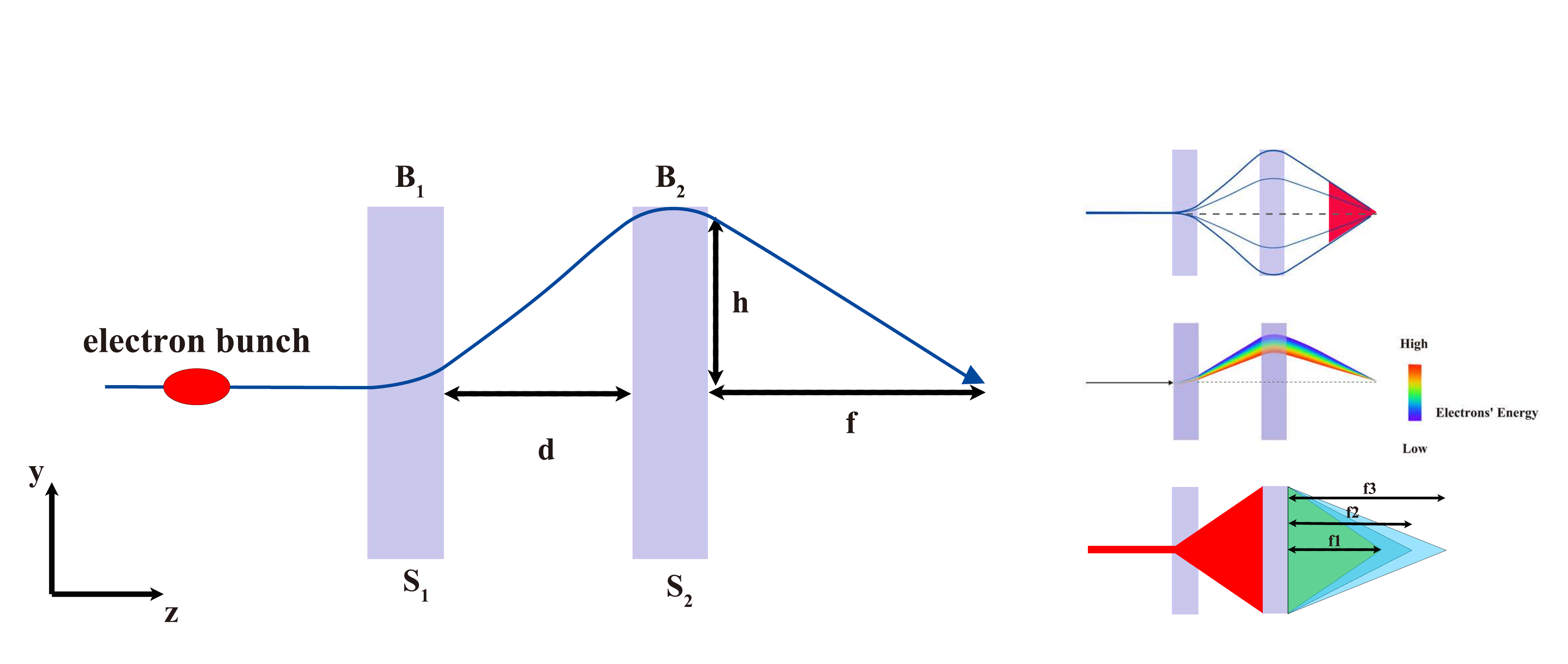}
    \caption{Schematic illustration of the proposed scheme.
    (a) The basic two-dipole layout and a typical electron beam trajectory.
    (b) A maximum dose at the target position is created by guiding electron beams through different angular trajectories via varying field strengths of both dipoles.
    (c) An electron beam with notable energy spread can be guided to target the same position.
    (d) The dose-peak depth is controlled by the strength ratio $k$.}
\end{figure*}

The basic layout of the proposed beam delivery system, is plotted in Figure 1(a), where the two dipoles with strengths and lengths denoted as $B_1$ and $B_2$, and $S_1$ and $S_2$, respectively, are separated by a distance of $d$, and they satisfy $|B_1S_1|<|B_2S_2|$ with $B_1$ and $B_2$ having opposite signs. We further consider an electron beam with an initial longitudinal momentum of $p_z$ and no divergence. The deflection angles of the beam through the first and second magnet ($\theta_1$ and $\theta_2$) can be roughly estimated as,
\begin{eqnarray}
    \theta_1 \approx \frac{S_1}{R_1},
    \\
    \theta_2 \approx \frac{S_2}{R_2},
\end{eqnarray}\label{theta}
where $R_{1,2}=\frac{p_z}{q B_{1,2}}$ represent the Larmor radius and the above equations are valid under the condition of $S_1, S_2\ll R$. The transverse displacement $h$ of the particle bunch after drifting a longitudinal distance $d$ can then be expressed as $h = {d}{\tan{\theta_1}} \approx {d}{\theta_1}$ based on thin-magnet and paraxial approximations. Since these two magnets have opposite signs, the bunch, after passing through the second magnet, finally get deflected back to the initial longitudinal axis and the distance between the crossing point and the rear side of the second magnet, denoted as $f$, can be expressed as,
\begin{eqnarray}
    f \approx \frac{h}{\theta_2 - \theta_1} = \frac{d}{\frac{R_1}{R_2}\frac{S_2}{S_1} - 1} = \frac{d}{\frac{B_2}{B_1}\frac{S_2}{S_1} - 1}\label{eq:two}.
\end{eqnarray}

Through Eq.~\ref{eq:two}, it is evident that for a given beam transport system with fixed magnetic lengths, the value of parameter $f$ is uniquely determined by the ratio $k$ of the strengths of two magnets, represented as $k = \frac{B_2}{B_1}$. This beam transport system offers the following three advantages in dose delivery:

Firstly, by scanning $B_1$ ($B_2$) from the positive (negative) maximum to the negative (positive) maximum while keeping their ratio $k$ constant, an electron bunch with equivalent energy can be guided towards the same position but along different angular directions. By aggregating these cases, a concentrated dose maximal can be achieved near the converging position where the tumor is located, as shown in Figure 1(b).

Secondly, Equation (\ref{eq:two}) is independent of the beam energy. Consequently, electrons with various energies passing through the system will ultimately converge at the same position. This characteristic implies that the system is resilient to beam's energy spread, which distinguishes it from other proposed schemes \cite{glinec2006radiotherapy, kokurewicz2021experimental, whitmore2021focused} (see Figure 1(c)).

Lastly, the beam converging position can be readily adjusted by modifying the strength relations of the two dipoles, providing exceptional flexibility for targeting tumors at different locations. Figure 1(d) shows three cases of different dose-peak depths by varying the ratio $k$.

Noting that, the above analytical model is quite simplified. When considering VHEE beams with initial divergence transported within a realistic magnetic field distribution, they will not be guided to converge to a single point but spread into a spot of finite size. However, this effect is significantly smaller than the multiple Coulomb scattering caused by the water medium when propagating deeply inside the phantom (i.e., 10-20 cm). Therefore, the three advantages previously mentioned remain valid. A detailed analysis of this point can be found in Section I and Figure S1 of the Supplementary Materials.

\subsection{Validation using Monte Carlo simulations}

\begin{figure*}
    \centering
    \includegraphics[width=1\linewidth]{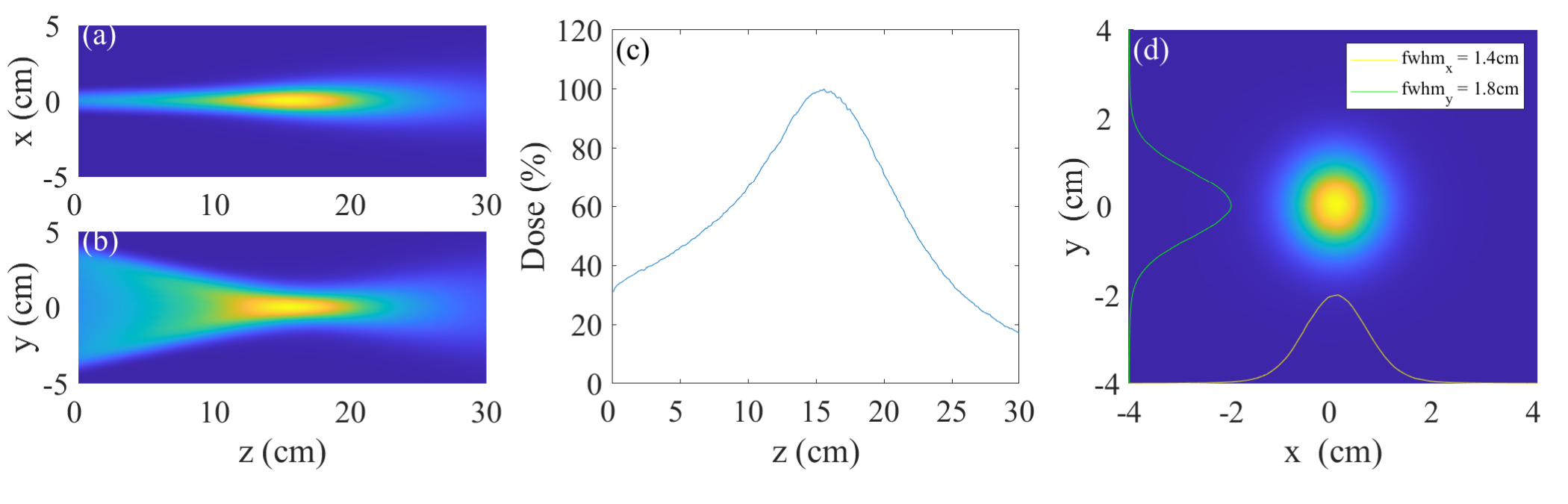}
    \caption{Monte Carlo simulation results of VHEE dose deposition in a water phantom using the proposed scheme. (a)/(b) 2D dose distribution along (x-z)/(y-z) plane, respectively. (c) On-axis dose distribution along the z axis, where an in-depth dose peak at 15 cm can be observed. (d) Transverse dose distribution at the $z=15$ cm plane, with FWHM sizes of 1.4cm/1.8cm in x and y directions, respectively.}
    \label{fig:enter-label}
\end{figure*}

In order to validate the proposed scheme, we have performed Monte Carlo simulations using the code TOPAS \cite{perl2012topas}. In these simulations, we defined the longitudinal direction as $z$, the transverse deflection direction as $y$ and the other transverse direction as $x$. The lengths of the two dipoles were set at $S_1$ = 10 cm and $S_2$ = 22 cm, with a gap size of 2 cm. The two magnets are separated by 30 cm. The magnetic field distributions, including the fringe fields, for both dipoles are calculated using three-dimensional (3D) electromagnetic simulation software CST (More detailed field distribution can be found in Section II and Figure S2 of the Supplementary Materials). The maximum strengths of the two magnets are chosen as 1.35 T, with $B_1$ and $B_2$ having opposite signs. A water phantom was positioned 30 cm behind the second magnet, with dimensions of 20 cm ($x$) $\times$ 20 cm ($y$) $\times$ 40 cm ($z$), and a voxel size of 1 mm in all directions. The entrance of the water phantom was defined as $z=0$. Considering a typical electron bunch originating from a standard LWFA, we set its initial distribution in both space and divergence space as Gaussian distribution, with RMS sizes as ${\sigma_x}$ = ${\sigma_y}$ = 4 $\mu$m and RMS divergences as ${\theta_x}$ = ${\theta_y}$ = 4 mrad, respectively. The bunch contained a peak energy of 200 MeV and a 5\% RMS energy spread, and was launched 15 cm ahead of the front surface of the first dipole. Based on this configuration, the total length of the beam delivery system from the source to the water phantom entrance is less than 1 meter. 

To achieve a maximum dose at a specific depth deeply inside the water phantom, we launched eleven simulations. In each simulation, $10^{6}$ particles were used and only the value of $B_1$ ($B_2$) was varied from the negative (positive) maximum to the positive (negative) maximum with a constant incremental step. After summing up all above cases, the resulted dose distribution were presented in Figure 2.   

Figures 2(a) and (b) present the central slice dose distribution in the $z-x$ and $z-y$ planes, respectively, where one can see the appearance of a maximum dose deposited deeply inside the water phantom. To further verify this finding, we present the longitudinal on-axis dose lineout in Figure 2(c). The dose peak is observed located at a depth of 15 cm, with the entrance dose only accounting for around 33\% of the peak dose. 
Furthermore, Figure 2(d) displays the transverse ($x-y$) dose distribution at the dose-peak depth, where the full width at half maximums (FWHMs) along the two dimensions were measured as FWHM$_y=1.8$ cm and FWHM$_x=1.4$ cm, respectively. These values are comparable to or even smaller than the dose size of a typical proton pencil beam at the same depth, suggesting that the combination of VHEE and this dose delivery system can greatly benefit accurate tumor targeting. 

To evaluate the system's tolerance to the bunch energy spread, we conducted four simulations with the same setup as previously described, but applying electron beams with different energy distributions. The results are presented in Figure 3(a-d). For simulations (a), (b), and (c), the energy spectra adhere to a Gaussian distribution with a central energy of 200 MeV and a RMS spread of 0.1\%, 5\%, and 20\%, respectively. In simulation (d), the energy spectrum is uniform, ranging from 100 MeV to 300 MeV. Despite significant disparities in energy spread, the system is shown to be capable of concentrating the dose peak with similar longitudinal and lateral dose distributions. Figure 3(e) provides a more detailed on-axis dose distribution along the z-axis for these four cases. The first three cases exhibit negligible differences, with only the fourth case demonstrating minor variation of less than 5\%, which confirms the robustness of the proposed scheme. We note that a dose peak may also be achieved by the asymmetric focusing using a series of quadrupoles \cite{whitmore2021focused}. However, to achieve a similar dose-peak depth (e.g.,$\sim$15 cm) with small entrance-to-peak dose ratio (e.g., $\sim$30\%), the required parameters for quadrupoles are much more challenging, and they cannot transport electrons with large energy dispersion as effectively as the presented system. 

\begin{figure*}[htbp]
	\centering
	\includegraphics[width=1.0\linewidth]{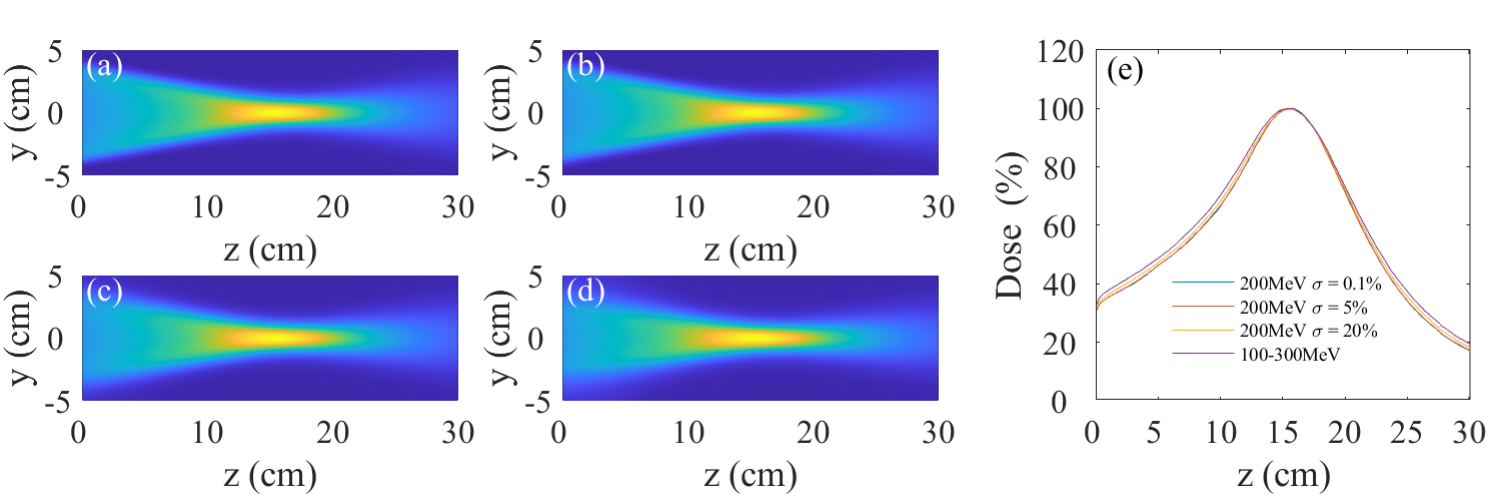}
	\caption{Relative dose depositions in a water phantom by Angular scanning VHEE beams with different energy spreads. (a)/(b)/(c) 2D dose distribution in the y-z plane by an VHEE beam with 0.1\%/5\%/20\% RMS relative energy spread. (d) 2D dose distribution by an electron beam with a flat-top energy spectrum ranging from 100MeV to 300MeV. (e) On-axis dose distribution of all the 4 cases. The mean energy of the electron beams for (a-d) is 200 MeV.}
\end{figure*}

\subsection{Controlling the dose peak position}

To ensure a precise and flexible dose delivery, a reliable system is necessary to enable the control over the position of dose deposition for targeting different types of tumors. Here we show that by simply modifying the two dipole field strengths, this requirement can be fulfilled. We consider a simple relation between them as $B_2 = k B_1 + b$, where $k$ and $b$ are controllable variables. This formula allows for the longitudinal (transverse) adjustment of the dose-peak position position by varying $k$ ($b$). A series of Monte Carlo simulations have been carried out to confirm this approach with the same setup as Figure 2 but different magnet strengths, and the results are presented in Figure 4. As shown in Figure 4(a), the dose-peak depth inside the water increases from 7 cm to 20 cm as $k$ decreases from 0.95 to 0.78, and the correlation between them is almost linear (see Figure 4(b)). It is noted that this increase in dose-peak depth may result in higher entrance doses owing to increased scattering of electron beams. Figure 4(c) illustrates the transverse dose lineouts along the $x$-axis at the dose-peak depth (15 cm) for different $b$ values. By setting $b$ from 0 T to 0.2 T, the dose-peak position can be laterally displaced by 3 cm in the corresponding directions along the $x$-axis, and the transverse displacement is also shown in linear dependence of the $b$ values (see Figure 4(d)). It is noted that during the transverse scanning (only varying $b$), the longitudinal dose-peak depth can also be shifted slightly. In our case, a lateral displacement of 3 cm results in an approximate 1 cm backward shift of the dose-peak depth. Nevertheless, this shift can be compensated by adjusting the parameter 
$k$. 

We also note that chromatic aberration due to large energy spread could potentially affect the lateral size of the dose-peak area during the transverse scanning of the dose-peak position. To address this concern, we conducted two sets of Monte Carlo simulations with the dose-peak transversely displaced by around 3 cm. Our findings indicate that the impact of chromatic aberration on the enlargement of the dose-peak size is insignificant compared to the contributions from the original divergence of LWFA beams (a few mrad) and multiple Coulomb scattering in water or human tissue. Detailed results are presented in Section III and Figures S3-S4 of the Supplementary Materials.

To enable the scanning in both transverse directions, the two dipole magnets (weighing less than 200 kg in total) can be mounted together onto a Gantry and rotated around the axis of the incoming beam. This configuration allows for a full scanning range covering a circular area with a diameter of 6 cm, sufficient for treating small to medium-sized tumors. 
Additionally, there are several ways to further increase the scan range. For instance, since the LWFA itself is only a few millimeters in size, we can easily displace the entire LWFA module (including the laser focusing optics) transversely by several centimeters.

\begin{figure*}
    \centering
    \includegraphics[width=1\linewidth]{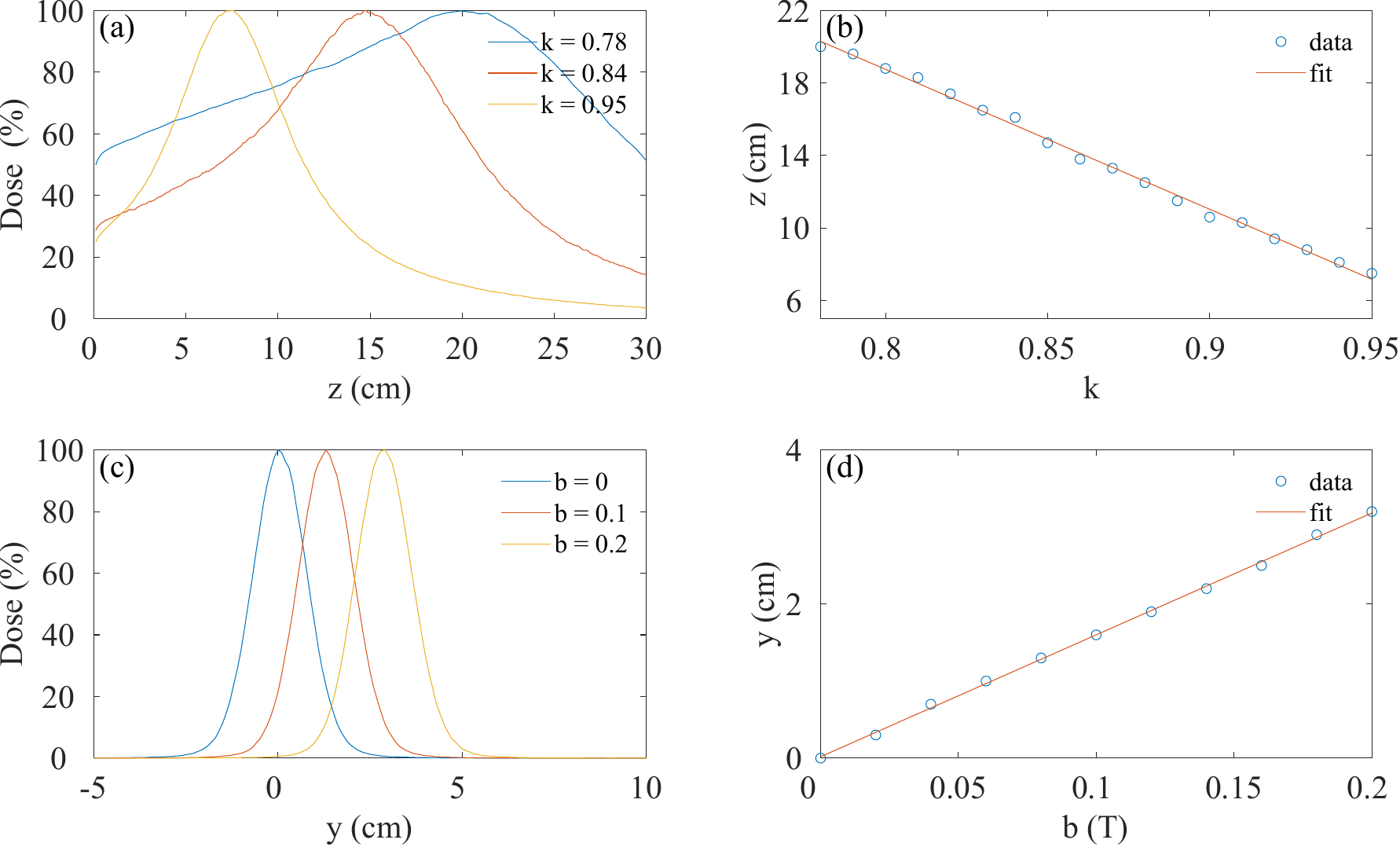}
    \caption{Dose-peak position control along longitudinal and transverse directions using the angular scanning scheme. (a) Relative longitudinal on-axis dose distribution with different parameter $k$. (b) Relationship between the depth of the dose-peak position and parameter $k$. (c) Transverse dose deposition lineout at $z=15$ cm in the x-z plane with b=0 (blue), b=0.1 (red) and b=0.2 (yellow). (d) Dependence of lateral displacement of the dose-peak position on parameter $b$. For (c)-(d), the blue circles are simulation results, and the red curve is a linear regression.}
    \label{fig:enter-label}
\end{figure*}

\subsection{Flat-top dose peak formation}

Similar to the commonly used Spread-Out Bragg Peak (SOBP) technique in proton therapy, which broadens the Bragg peak by linearly superposing different proton energy, we can also generate a Spread-Out Electron Peak (SOEP) by summing up doses of different dose-peak depths assigned with specific weights. This concept was introduced in \cite{whitmore2021focused}, and in our case, it can be accomplished by simply modifying the value of the parameter $k$. Figure 5 illustrates two simulation examples of the SOEP for a 200 MeV VHEE beam. In Figure 5(a), a dose plateau extending from a depth of 8 cm to 13 cm is presented, and Figure 5(b) depicts a plateau spanning from a depth of 8 cm to 18 cm. The shaded green region within each figure denotes the desired depth range, where the achieved dose uniformity is quite good, with the relative standard deviations of $\frac{\sigma D_{on-axis}}{D_{on-axis}}=0.52\%$ in Figure 5(a) and $\frac{\sigma D_{on-axis}}{D_{on-axis}}=0.45\%$ in Figure 5(b). Additionally, the grey curves in each figure represent the on-axis doses of different dose-peak depths after applying different weights. 

\begin{figure*}
    \centering
    \includegraphics[width=1\linewidth]{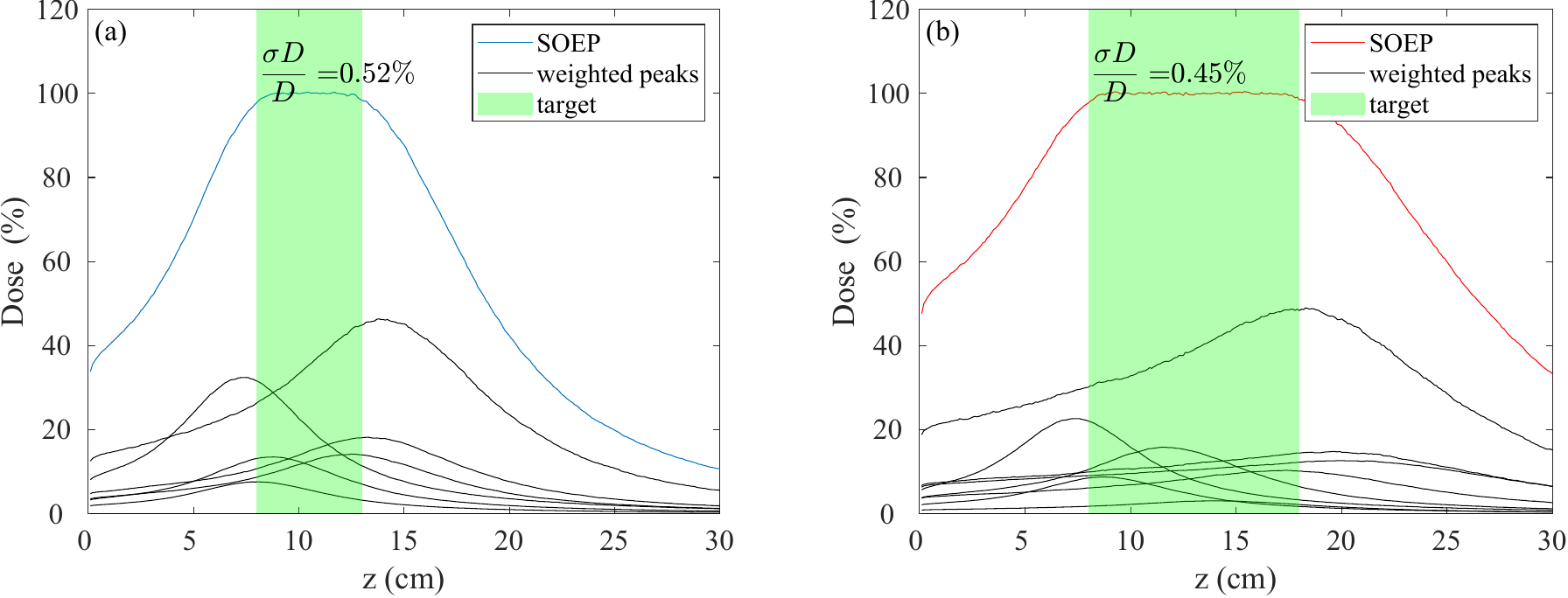}
    \caption{Simulation results of achieving Spread-Out Electron Beam (SOEB) by angular scanning scheme. (a) Accumulating 6 weighted electron dose profiles resultes in a flat region with a length of 5 cm at a depth ranging between 8-13 cm and (b) Accumulating 8 weighted electron dose profiles resultes in a flat region with a length of 10 cm at a depth ranging between 8-18 cm.}
    \label{fig:enter-label}
\end{figure*}

\section{Conclusion}

In this article, we propose a simple dose delivery scheme that utilizes only two dipole magnets to effectively target deep-seated tumors with a concentrated dose peak in a very compact configuration. By adjusting the correlations between these two magnets, we can precisely position the dose peak in both longitudinal and lateral directions. An additional advantage of this system is its resilience to broadband energy spectra, which makes it particularly appealing for the use in LWFA-based VHEE radiotherapy. In the future, we plan to incorporate the laser acceleration module and the proposed dose delivery system into a rotational small Gantry, equipped with advanced diagnostic tools such as femtosecond high-energy electron probe \cite{wan2022direct,wan2023femtosecond,wan2024real}, advanced plasma shaping tools \cite{seemann2023refractive,seemann2024laser} and machine learning functions \cite{Jalas2021Bayesian,Irshad2024Pareto}, which will optimize the LWFA performance and enable a dose delivery with large scanning rangle, precise tumor conformity and high operational flexibility, crucial for the development of LWFA-based VHEE radiotherapy devices towards clinical applications.

\section {Acknowledgement}
This work was supported by the Strategic Priority Research Program of the Chinese Academy of Sciences (Grants No. XDB0530000, No. XDB0530100 and No. XDB0530200) to W.L. and Y.W., National Natural Science Foundation of China (Grants No. 11991071 to W.L. and No. 12405169 to S.L.), Science Fund Program for Distinguished Young Scholars of the National Natural Science Foundation of China (Overseas) to Y.W., Discipline Construction Foundation of “Double World-class Project” to W.L. and J.H., Key Scientific Research Projects of Henan Provincial Colleges and Universities No. 25ZX002 to Y.W.. The simulations were performed at National Supercomputing Center in Zhengzhou and Center of High performance computing, Tsinghua University.

Bing Zhou and Zhiyuan Guo contributed equally to this work.

\providecommand{\noopsort}[1]{}\providecommand{\singleletter}[1]{#1}%

\end{document}